\documentclass[11pt,preprint]{aastex}

\newcommand{\diff}{\mbox{${\rm d}$}}

\newcommand{\feh}{\mbox{\rm [{\rm Fe}/{\rm H}]}}
\newcommand{\mh}{\mbox{\rm [{\rm M}/{\rm H}]}}
\newcommand{\Msun}{\mbox{$M_{\odot}$}}
\newcommand{\Teff}{\mbox{$T_{\rm eff}$}}

\newcommand{\beq}{\begin{equation}}
\newcommand{\eeq}{\end{equation}}
\newcommand{\beqa}{\begin{eqnarray}}
\newcommand{\eeqa}{\end{eqnarray}}

\slugcomment{To appear in PASP May 2008 issue \\
(see http://www.journals.uchicago.edu)}

\begin{document}


\title{Revised bolometric corrections and interstellar extinction 
coefficients for the ACS and WFPC2 photometric systems}


\author{L\'eo Girardi}
\affil{Osservatorio Astronomico di Padova -- INAF, Padova, Italy}
\author{Julianne Dalcanton, Benjamin Williams}
\affil{University of Washington, Seattle}
\author{Roelof de Jong}
\affil{Space Telescope Science Institute, Baltimore}
\author{Carme Gallart, Matteo Monelli}
\affil{Instituto de Astrofisica de Canarias, La Laguna, Tenerife, Spain}
\author{Martin A.T. Groenewegen}
\affil{Katholieke Universiteit Leuven, Belgium}
\author{Jon A. Holtzman}
\affil{New Mexico State University, Las Cruces NM}
\author{Knut A.G. Olsen}
\affil{Gemini Science Center, NOAO, Tucson}
\author{Anil C. Seth}
\affil{Harvard Center for Astrophysics, Boston}
\author{Daniel R. Weisz}
\affil{University of Minnesota, Minneapolis}
\author{(for the ANGST/ANGRRR collaboration)}




\begin{abstract}
We present extensive tables of bolometric corrections and interstellar
extinction coefficients for the WFPC2 and ACS (both WFC and HRC)
photometric systems. They are derived from synthetic photometry
applied to a database of spectral energy distributions covering a
large range of effective temperatures, surface gravity, and metal
content. Carbon stars are also considered. The zero-points take into
consideration the new high-accuracy Vega fluxes from Bohlin. These
tables are employed to transform Padova isochrones into WFPC2 and ACS
photometric systems using interstellar extinction coefficients on a
star-to-star basis. All data are available either in tabular form or
via an interactive web interface in the case of the
isochrones. Preliminary tables for the WFC3 camera are also included
in the database.
\end{abstract}





\section{Introduction}

The Advanced Camera for Surveys (ACS) aboard the Hubble Space
Telescope (HST) has provided some of the deepest photometric optical
images ever taken.  At the same time, the Wide Field and Planetary
Camera 2 (WFPC2) camera aboard HST has a rich photometric data archive
dating back more than a decade.  Both cameras have taken many images
of galaxies within $\sim$5 Mpc, providing reliable photometry for
individual stars within these nearby galaxies.  This resolved stellar
photometry contains detailed information about the history of star
formation and chemical enrichment.  Extracting this information
requires stellar evolution isochrones that track the photometric
properties of stars of different ages and metallicities.  Only with
precise models can these rich samples of stellar photometry provided
by ACS and WFPC2 be translated into coherent histories of star
formation and chemical evolution.

Given the extreme importance of the stellar photometric data provided
by ACS and WFPC2, many authors have described the conversion from
theoretical stellar models into these systems. For the WFPC2 system,
transformations and isochrones have been provided by Holtzman et
al. (1995), Chiosi et al. (1997), Salasnich et al. (2000), Girardi et
al. (2002), Lejeune (2002), and Dotter et al. (2007). For ACS, BC
tables and isochrones have been provided at
\url{http://pleiadi.oapd.inaf.it} since 2003, as an addition to the
Girardi et al. (2002) database, and later by Bedin et al. (2005) and
Dotter et al. (2007).

With the present paper, our goal is to provide revised and homogeneous
tables of BCs and interstellar extinction coefficients in the ACS and
WFPC2 systems, updating as much as possible the files distributed by
Girardi et al. (2002) in precedence. The motivation for an overall
revision of the BC tables reside mainly on the definition of better
spectrophotometric standards by Bohlin and collaborators (Bohlin 2007,
and references therein), and more complete libraries of stellar
spectra, whereas filter transmission curves for these instruments
remain essentially the same as in the original papers by the
instrument teams.  A consistent set of interstellar extinction
coefficients for these systems would be particularly important since
the unique resolution of HST allows useful optical photometry to be
obtained even for heavily reddened regions (with say $A_V\ga3$; see
e.g. Recio Blanco et al. 2005), for which interstellar extinction
coefficients being used so far may not be accurate enough.

The BC and interstellar extinction tables will be used to transform
theoretical isochrones to these systems, and will be extensively
applied in the interpretation of ANGST/ANGRRR data\footnote{ANGST (ACS
Nearby Galaxy Survey) is a Treasury HST proposal for the imaging of a
complete sample of galaxies up to a distance of 3.5 Mpc. Together with
archival data re-reduced by the ANGRRR (Archive of Nearby Galaxies -
Reduce, Recycle, Reuse) collaboration, it will allow the measurement
of their star formation histories with a resolution of 0.3~dex in
log(age), up to the oldest ages.}.

We proceed as follows: Sect.~\ref{sec_bc} describes the steps involved
in the derivation of bolometric corrections via synthetic photometry,
from the assembly of a spectral library to the definition of
zeropoints. Sect.~\ref{sec_extinction} describes the interstellar
extinction coefficients.  The results are used in Sect.~\ref{sec_isoc}
to convert theoretical isochrones into the ACS and WFPC2 systems,
which are made publicly available. They are briefly illustrated in
Sect.~\ref{sec_data}, by means of a few comparisons with observations.

\section{Bolometric corrections}
\label{sec_bc}

Given the spectral fluxes at the stellar surface, $F_\lambda$,
bolometric corrections for any set of filter transmission curves
$S_\lambda$\footnote{Throughout this paper, $S_\lambda$ is the total
system throughput, i.e. including the telescope, camera, and filters.} 
are given by (see Girardi et al. 2002 for details)
	\beqa
BC_{S_\lambda} & = & M_{\rm bol, \odot} 
	- 2.5\,\log \left[ 
		4\pi (10\,{\rm pc})^2 F_{\rm bol}/L_\odot
		\right] \label{eq_bcfinal} 
	\\ \nonumber
	&& + 2.5\,\log\left(
	\frac { \int_{\lambda_1}^{\lambda_2} 
		\lambda F_\lambda \, 10^{-0.4A_\lambda} 
		S_\lambda \diff\lambda }
		{ \int_{\lambda_1}^{\lambda_2} 
		\lambda f^0_\lambda S_\lambda \diff\lambda } 
	\right)
	- m_{S_\lambda}^0
	\eeqa
where $f^0_\lambda$ represents a reference spectrum (incident on
the Earth's atmosphere) that produces a known apparent magnitude
$m^0_{S_\lambda}$, $F_{\rm bol} = \int_0^\infty F_\lambda
\diff\lambda = \sigma T_{\rm eff}^4 $ is the total emerging flux 
at the stellar surface, and $A_\lambda$ is the interstellar extinction
curve in magnitude units. Once $BC_{S_\lambda}$ are computed, stellar
absolute magnitudes follow from
	\beq
M_{S_\lambda} = M_{\rm bol} - BC_{S_\lambda} \,\,\, ,
\label{eq_absolmag}
	\eeq
where
	\beqa
M_{\rm bol} & = & M_{\rm bol, \odot} - 2.5\, \log(L/L_\odot)
	\label{eq_mbol}
	 \\ \nonumber
	 & = & M_{\rm bol, \odot} - 2.5\, \log(
		4\pi R^2 F_{\rm bol} /L_\odot) \,\,\,\, .
	\eeqa
As in Girardi et al. (2002), we adopt $M_{\rm bol,
\odot}=4.77$, and $L_\odot=3.844\times10^{33}\,{\rm erg\,s^{-1}}$
(Bahcall et al.\ 1995).

Notice that Eq.~\ref{eq_bcfinal} uses photon count integration instead
of energy integration, as appropriate to most modern photometric
systems that use photon-counting detectors instead of
energy-amplifiers. The equation also includes the interstellar
extinction curve $A_\lambda$, that will be discussed later in
Sect.~\ref{sec_extinction}. $A_\lambda=0$ is assumed for the moment.

\subsection{A library of spectral fluxes}
\label{sec_spectra}

Girardi et al. (2002) assembled a large library of spectral fluxes,
covering wide ranges of initial metallicities (\mh\ from $-2.5$ to
$+0.5$), \Teff\ (from 500 to 50\,000~K), and $\log g$ (from $-2$ to
$5$). This grid is wide enough to cover most of the stellar types that
constitute the bulk of observed samples in resolved galaxies. The grid
is composed of ATLAS9 ``NOVER'' spectra from Castelli et al (1997, see
also Bessell et al. 1998) for most of the stellar types, Allard et
al. (2000a) for M, L and T dwarfs, Fluks et al. (1994) empirical
spectra for M giants, and finally pure blackbody spectra for stars
exceeding 50\,000~K.  Since then, we have updated part of this library,
namely:

1) As part of the TRILEGAL project (Girardi et al. 2005), we have
included spectra for DA white dwarfs from Finley et al. (1997) and
Homeier et al. (1998) for \Teff\ between 100\,000 and 5\,000~K.

2) We now use Castelli \& Kurucz (2003) ATLAS9 ``ODFNEW'' spectra,
which incorporate several corrections to the atomic and molecular
data, especially regarding the molecular lines of CN, OH, SiO, H$_2$O
and TiO, and the inclusion of HI--H$^+$ and HI--H$^+$ quasi-molecular
absorption in the UV. These new models are considerably better than
the previous NOVER ones, especially in the UV region of the spectrum,
and over most of the wavelength range for $\Teff$ between $\sim4500$
and $3500$~K (see figs. 1 to 3 in Castelli
\& Kurucz 2003).

3) We have also incorporated in the library the Loidl et al. (2001)
spectral fluxes of solar-metallicity carbon (C-) stars derived from
static model atmospheres. They are computed for \Teff\ in the range
from 2600 to 3600~K, and for C/O$=1.1$ and 1.4. The C/O$=1.1$ spectra
are now used for all C/O$>1$ models, replacing the Fluks et al. (1994)
spectra of M giants which were previously (and improperly) used for
all cool giants including C-rich ones.

Regarding this latter aspect, it is to be noticed that all Padova
isochrones published since 2001 (Marigo \& Girardi 2001; Cioni et
al. 2006; Marigo et al. 2008) do take into consideration the third
dredge-up events and hence the conversion to a C-type phase during the
TP-AGB evolution. This phase, characterised by the surface C/O$>1$,
has to be represented with either proper C-rich spectra, or
alternatively, with empirical $\Teff$-colour relations as in Marigo et
al. (2003). The use of Loidl et al. (2001) spectra frees us definitely
from using empirical relations for C stars. Note that such empirical
relations would not even be available for the ACS and WFPC2 filters we
are interested in. In fact, empirical $\Teff$-colour relations for C
stars are usually limited to red and near-IR filters, and mainly to
$IJHK$. 

Although the Loidl et al. (2001) models are quite limited in their
coverage of the parameter space of C stars (i.e. [M/H], C/O, \Teff,
and $\log g$) they represent a useful first approach toward obtaining
realistic colours for them. Work is in progress to extend
significantly the parameter space of such models (Aringer et al., in
preparation).

\subsection{Filters and zeropoints}
\label{sec_zeropoints}

As can readily be seen in Eq.~\ref{eq_bcfinal}, in the formalism
adopted here, the photometric zeropoints are fully determined by a
reference spectral energy distribution $f^0_\lambda$, which provides a
given set of reference magnitudes $m_{S_\lambda}^0$.

In this work, we adopt the latest Vega spectral energy distribution
from Bohlin (2007) as $f^0_\lambda$. This Vega spectrum represents a
great improvement over previous ones distributed and used in the
past. It is the result of an effective effort using STIS to provide a
set of stellar spectrophotometric standards with an absolute
calibration to better than 1 percent at near-UV to blue
wavelengths. The best-fitting ATLAS9 model for Vega in this spectral
range is then used to complement its spectral energy distribution at
red and near-IR wavelengths. From Bohlin (2007), one can conclude the
new Vega spectrum is likely accurate to about 1--2 percent over the
complete 3200--10000~\AA\ interval.

Having adopted Bohlin's (2007) Vega spectrum, we have just to specify
the Vega magnitudes in ACS and WFPC2 filters to completely define our
zeropoints. Following Holtzman et al.'s (1995) definition of the WFPC2
synthetic system, we define that Vega has the same magnitudes in WFPC2
filters as in the Johnson-Cousins filters that are closest to them in
wavelength. This choice means a magnitude of 0.02 for all filters from
F336W to F450W, 0.03 for F555W and F569W, 0.035 from F606W to F850LP,
but for F702W which has magnitude 0.039. For filters blueward of
F300W, Vega magnitudes are assumed to be zero.

Then, for ACS we adopt the same definition as Sirianni et al. (2005),
that Vega has zero magnitudes in all ACS filters. For a question of
completness, we provide tables also for the AB and ST magnitude
systems.

These are of course synthetic definitions of WFPC2 and ACS systems,
that may present small systematic offsets with respect to the real
systems (such as the flight WFPC2 system defined by Holtzman et
al. 1995) to which observations are commonly transformed.

Finally, the filter transmission curves $S_\lambda$ -- including the
telescope and camera throughputs -- for WFPC2 are created with
SYNPHOT/STSDAS v3.7 using the most up-to-date throughput and
sensitivity reference files available at January 15, 2008.  For
WFPC2/F170W and WFPC2/F218W, instead, they are taken from Holtzman et
al. (1995). All of the wide filters are considered.  ACS filter
throughputs are from Sirianni et al. (2005). They are almost identical
to those provided by SYNPHOT/STSDAS v3.7.

\subsection{Behaviour of bolometric corrections}
\label{sec_bolom}

\begin{figure}[h]
\plottwo{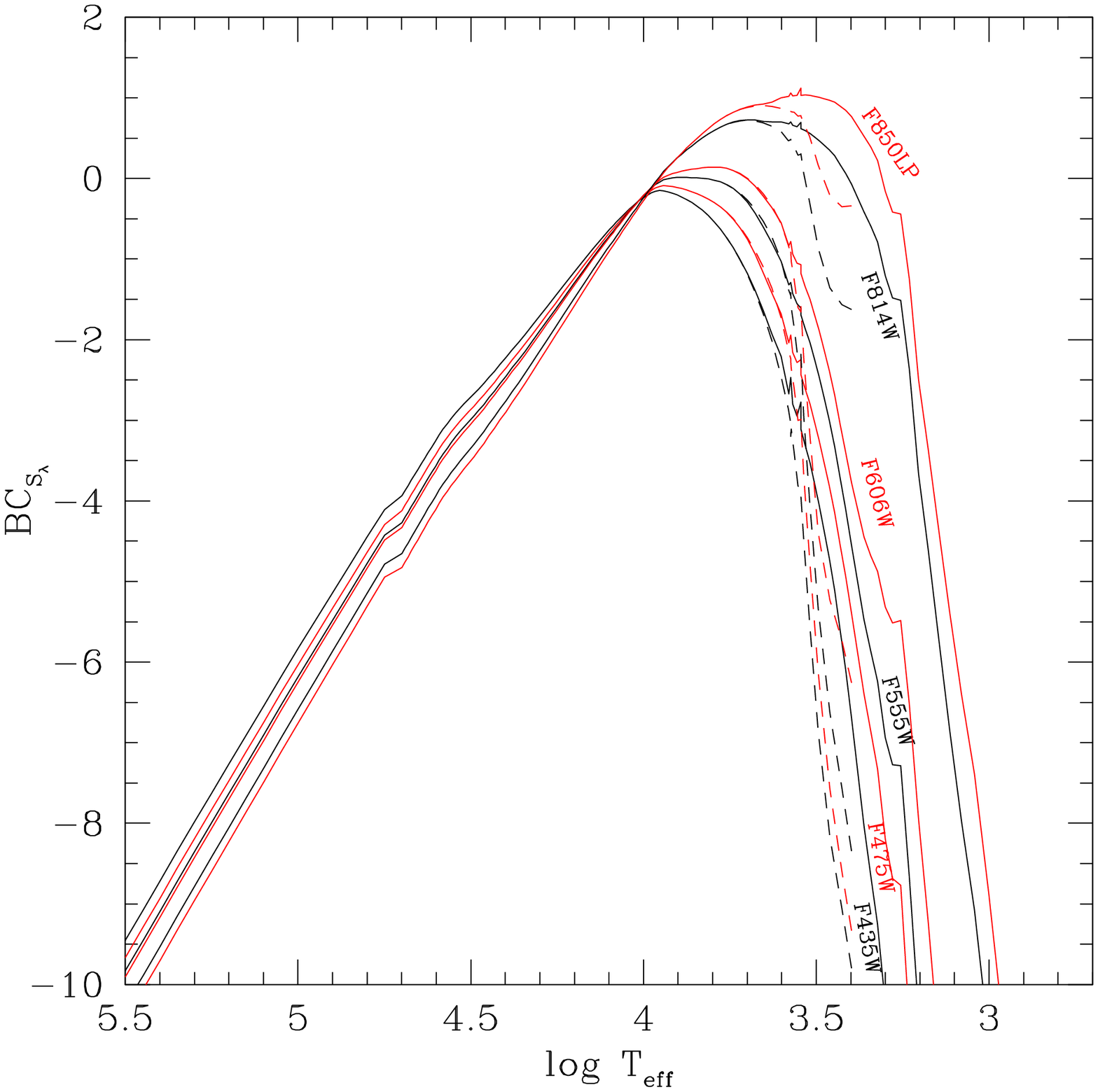}{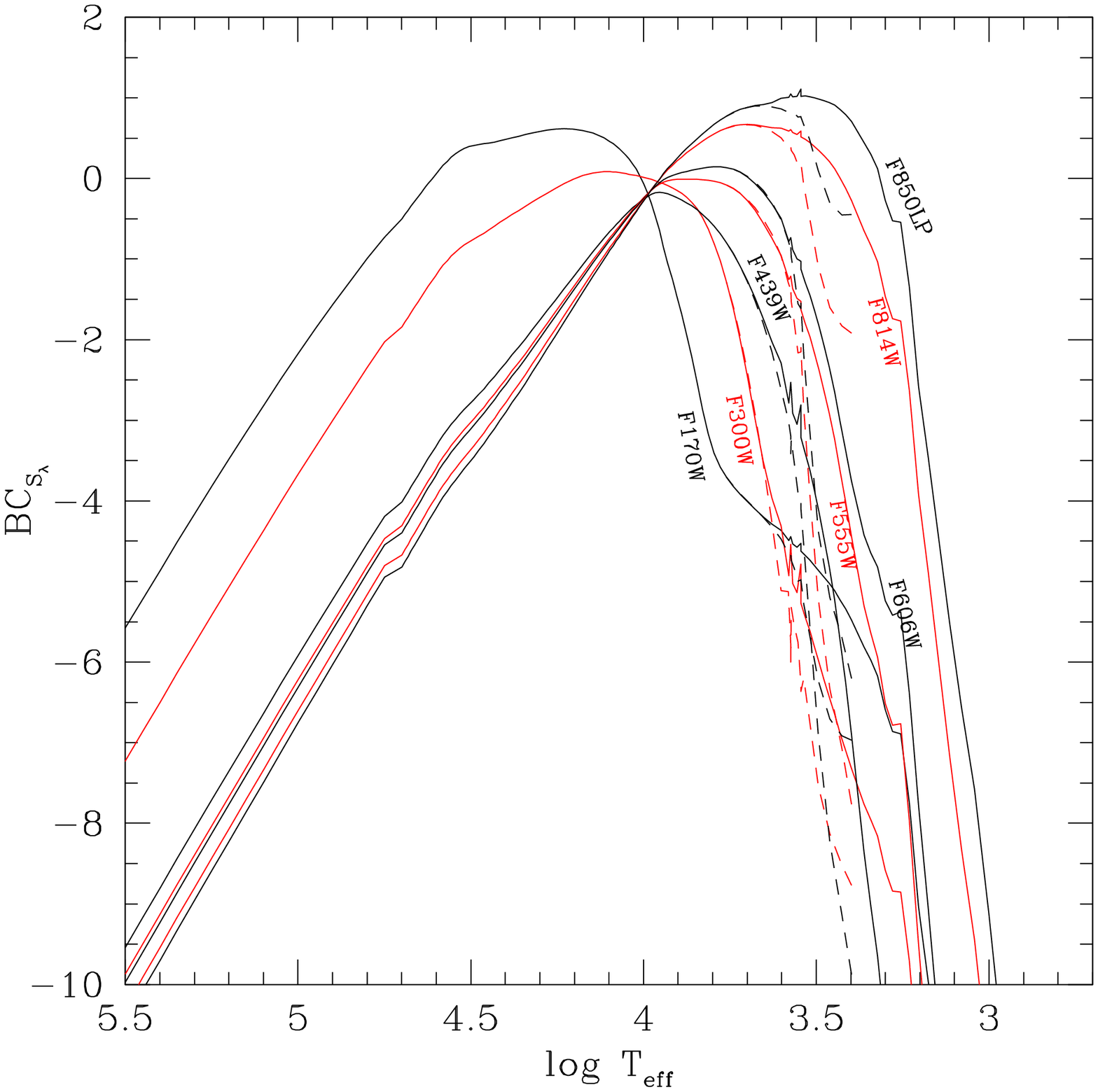}
\caption{$BC_{S_\lambda}$ as a function of \Teff, for 
both dwarfs (continuous lines) and giants (dashed lines), for some of
the ACS/WFC and WFPC2 filters. }
\label{fig_bc}
\end{figure}

\placefigure{fig_bc}

Figure~\ref{fig_bc} shows how BCs depend on \Teff\ for a series of ACS
and WFPC2 filters. In practice, we illustrate it for a sequence of
dwarfs of $\log g=5$, and for a sequence of subgiants and giants (of M
type at the coolest \Teff) following the relation
$\Teff=3250+500\,\log g$ (in c.g.s. units), which roughly describes the
position of the Hayashi track in the HR diagram of low-mass stars with
near-solar metallicity. The central part of the
\Teff\ interval reflects the behaviour of ATLAS9 synthetic
spectra. The small discontinuities in the figure occur as we change
from one spectral library to another, namely:
\begin{itemize}
\item The wiggles at $\log\Teff\simeq3.54$ occur when we pass 
from ATLAS9 to Fluks et al. (1994) spectra for M giants, and to Allard
et al. (2000a) spectra for M dwarfs.
\item The step at $\log\Teff\simeq3.27$ occurs when we change between the 
``AMES'' and ``AMES-dusty'' among Allard et al. (2000a) spectra. The
latter include dust formation in the outer atmosphere, and are
computed for solar-metallicity only (see Chabrier et al. 2000; Allard
et al. 2000b, 2001 for more details).
\item A second, less pronounced step at $\log\Teff\simeq4.75$
occurs when we start using blackbody spectra instead of ATLAS9.
\end{itemize}

Among these discontinuities, the most important are the wiggles at
$\log\Teff\simeq3.54$ because they occur at \Teff\ in which many stars
of Local Group galaxies are routinely observed in optical to near-IR
passbands. In order to limit the effect of such wiggles in the final
isochrones, we adopt a smooth transition between the different sources
of BC, for both dwarfs and giants, over the $\log\Teff$ interval from
3.542 and 3.590. This is the reason why isochrones using these
transformations (see the Fig.~\ref{fig_isocav} later for an example)
do not present significant discontinuities.

It is also interesting to note, in Fig.~\ref{fig_bc}, the anomalous
behaviour of F170W in WFPC2 which for cool stars provides brighter
magnitudes than redder filters. This behaviour is a consequence of its
red leak (see also Holtzman et al. 1995). A similar effect is also
present, but much less pronounced, in the F300W filter.

\section{Interstellar extinction coefficients}
\label{sec_extinction}

As already mentioned, Eq.~\ref{eq_bcfinal} can be used together with
an interstellar extintion curve, $A_\lambda$, to derive bolometric
corrections already including the effect of interstellar
extinction. Alternatively, we can derive relative interstellar
extinction coefficients, $A_{S_\lambda}/A_{V}$, by simply using
\beq
A_{S_\lambda}/A_V = [ BC_{S_\lambda}(0)-BC_{S_\lambda}(A_V) ] 
	/ A_V
\label{eq_extcoef}
\eeq
where the 0 stands for the $A_\lambda=0$ case.

The interstellar extinction curve $A_\lambda$ we use in
Eq.~\ref{eq_bcfinal} is taken from Cardelli et al. (1989) with
$R_V=A_V/E_{B\!-\!V}=3.1$, and integrated with the O'Donnell (1994)
correction for $909<\lambda/{\rm \AA}<3030$.  Other extinction curves
and $R_V$ values can be easily implemented whenever necessary
(e.g. Vanhollebeke et al. 2008). Our calculations are essentially
identical to those by Holtzman et al. (1995) and Sirianni et
al. (2005), but are now performed for a more extended range of stellar
parameters and extinction values.

Before proceeding, it is interesting to note that the two parameters
in Cardelli et al. (1989) interstellar extinction curve, $A_V$ and
$R_V$, refer to the Johnson's $B$ and $V$ bands. However, this does
not imply that synthetic realisations of the Johnson system will
recover the same values of $R_V$ and $A_V$ when the extinction curve
is applied. As examples, we find that for a G2V star the Bessell
(1990) version of the Johnson system provide $A_{V,{\rm
Bessell}}=1.006\,A_V$ and $R_{V,{\rm Bessell}}=3.457$, whereas
Ma\'{\i}z Apell\'aniz (2006) version gives essentially the same
numbers, with differences appearing just at the fifth decimal. These
numbers change somewhat as a function of spectral type and $A_V$
itself (as will be illustrated below for HST systems). Therefore,
$A_V$ should be considered as just ``the parameter that describes the
total amount of interstellar extinction in Cardelli et al. (1989)
law'', rather than a precise physical measure of the $V$-band
extinction. A similar comment applies also to $R_V$: in the context of
this paper it is ``the parameter that describes the shape of the
interstellar extinction curve'', rather than the total-to-selective
interstellar extinction of Johnson-like systems.

\begin{figure}[h]
\epsscale{0.5}
\plotone{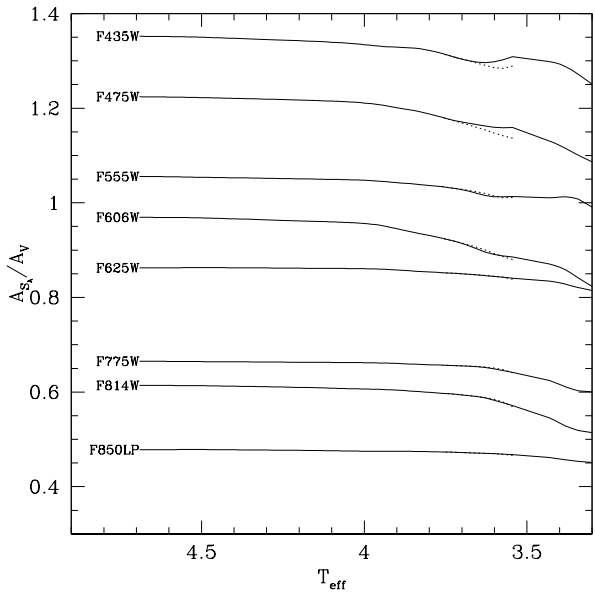}
\epsscale{1.0}
\caption{The variation of $A_{S_\lambda}/A_V$ as a function of \Teff, for 
both dwarfs (continuous lines) and giants (dashed lines), for some of
the ACS/WFC filters. }
\label{fig_avteff}
\end{figure}

\placefigure{fig_avteff}

The interstellar extinction coefficients derived from
Eq.~\ref{eq_extcoef} depend on the $F_\lambda$ under consideration,
and hence on parameters such as \Teff, $\log g$, and [M/H].
Figure~\ref{fig_avteff} illustrates the variation of
$A_{S_\lambda}/A_V$ as a function of $\Teff$, for several ACS/WFC
filters, and for both dwarfs (continuous lines) and giants (dashed
ones). Giants are again defined by the relation $\Teff=3250+500\,\log
g$.  The figure shows that $A_{S_\lambda}/A_V$ can change by as much
as $\sim0.3$ between very cool and very hot stars, especially for the
bluest (F435W, F475W) and widest (F606W, F814W) ACS filters. This
variation will inevitably change the morphology of a reddened
isochrone in the CMD, as compared to the unreddened case. This is
actually the main reason for taking star-to-star interstellar
extinction into consideration.

$A_{S_\lambda}/A_V$ variations are more sizeable in the regime of cool
temperatures and blue filters, which however is not the regime
targeted by most HST observations of stellar populations. The most
relevant \Teff\ interval for $A_{S_\lambda}/A_V$ variations is instead
between $\sim4000$ and 15000~K, which comprehends the main sequence
turn-off, horizontal branch, and tip of the red giant branch (TRGB) of
old metal-poor populations. Inside this interval, $A_{S_\lambda}/A_V$
variations amount to $\sim0.15$, and hence become potentially
important -- causing $\sim0.05$~mag effects in colours and magnitudes
-- at $A_V$ higher than $\sim3$~mag.

Another effect that can be accessed by our calculations is the
``Forbes effect'', or interstellar nonlinear heterochromatic
extinction (Forbes 1842). It consists of the variation of the relative
extinction $A_{S_\lambda}/A_V$ as the total extinction $A_V$
increases, and results from extinction being more effective for the
spectral regions with the higher flux. The $A_{S_\lambda}/A_V$
variations are expected to occur whenever the flux $F_\lambda$ varies
within the wavelength interval of a passband, and they are obviously
higher for wider passbands and at higher total extinction. This effect
was originally conceived to describe atmospheric extinction, but has
been shown to apply also for interstellar extinction by Grebel \&
Roberts (1995).

\begin{figure}[h]
\epsscale{0.5}
\plotone{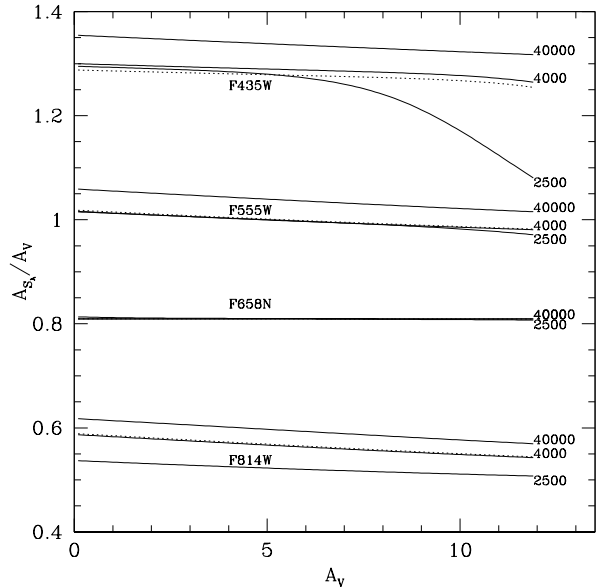}
\epsscale{1.0}
\caption{The Forbes effect for a few ACS filters, for dwarfs of 
$\Teff=2500$~K, $4000$~K and $40000$~K (continuous lines), and giants
with $\Teff=4000$~K (dashed lines). The \Teff\ values in K are
indicated at the right extremity of their curves.}
\label{fig_forbes}
\end{figure}

\placefigure{fig_forbes}

Figure~\ref{fig_forbes} illustrates the Forbes effect for some of the
ACS and WFPC2 passbands. It plots $A_{S_\lambda}/A_V$ as a function of
$A_V$ (solid lines), for a few values of \Teff\ and for the $0<A_V<12$
interval. It can be easily noticed that $A_{S_\lambda}/A_V$ decreases
with $A_V$, and that this decrease is more marked for wider
filters. In fact, $A_{S_\lambda}/A_V$ is almost constant for the
narrow F660N filter. Among the wide filters, the hottest spectra
($\sim40000$~K) present the most moderate variations in
$A_{S_\lambda}/A_V$, which is of approximately $0.005\,A_V^{-1}$ for
the most widely-used filters F555W and F814W.

\begin{figure}[h]
\epsscale{0.5}
\plotone{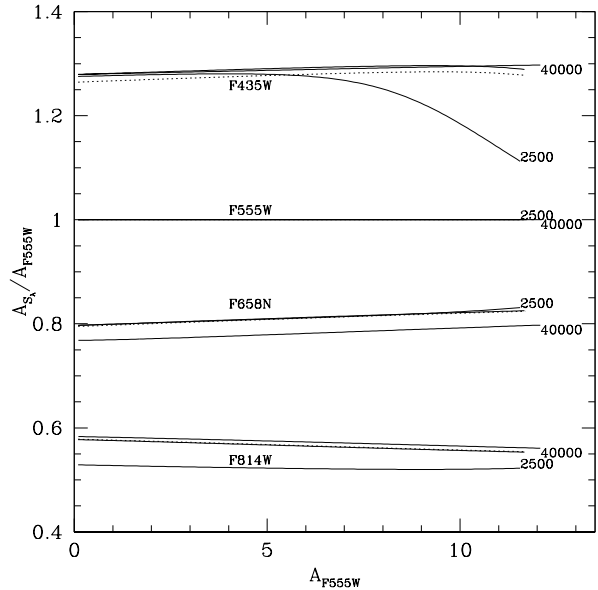}
\epsscale{1.0}
\caption{The same as Fig.~\ref{fig_forbes}, but now as a function of
$A_{\rm F555W}$, in the ACS/WFC system.}
\label{fig_forbes2}
\end{figure}

\placefigure{fig_forbes2}

Figure~\ref{fig_forbes2} instead plots $A_{S_\lambda}/A_{\rm F555W}$
as a function of $A_{\rm F555W}$, in the ACS/WFC system. This changes
somewhat the description of the Forbes effect, which becomes
increasing for F435W and F628N, and decreasing for F814W. Actually,
the most striking manifestation of the Forbes effect is expected to be
on the stellar colours, and this will be independent of the wavelength
taken as a reference for the total extinction.

Finally, we note that for stars cooler than 3000~K and for the bluest
ACS/WFC filters (for instance F435W) the Forbes effect becomes
dramatic already at $A_V\ga4$. These cases however are of little
practical interest since observations of heavily reddened cool stars
would hardly be done with such blue filters.

\begin{figure}[h]
\epsscale{0.5}
\plotone{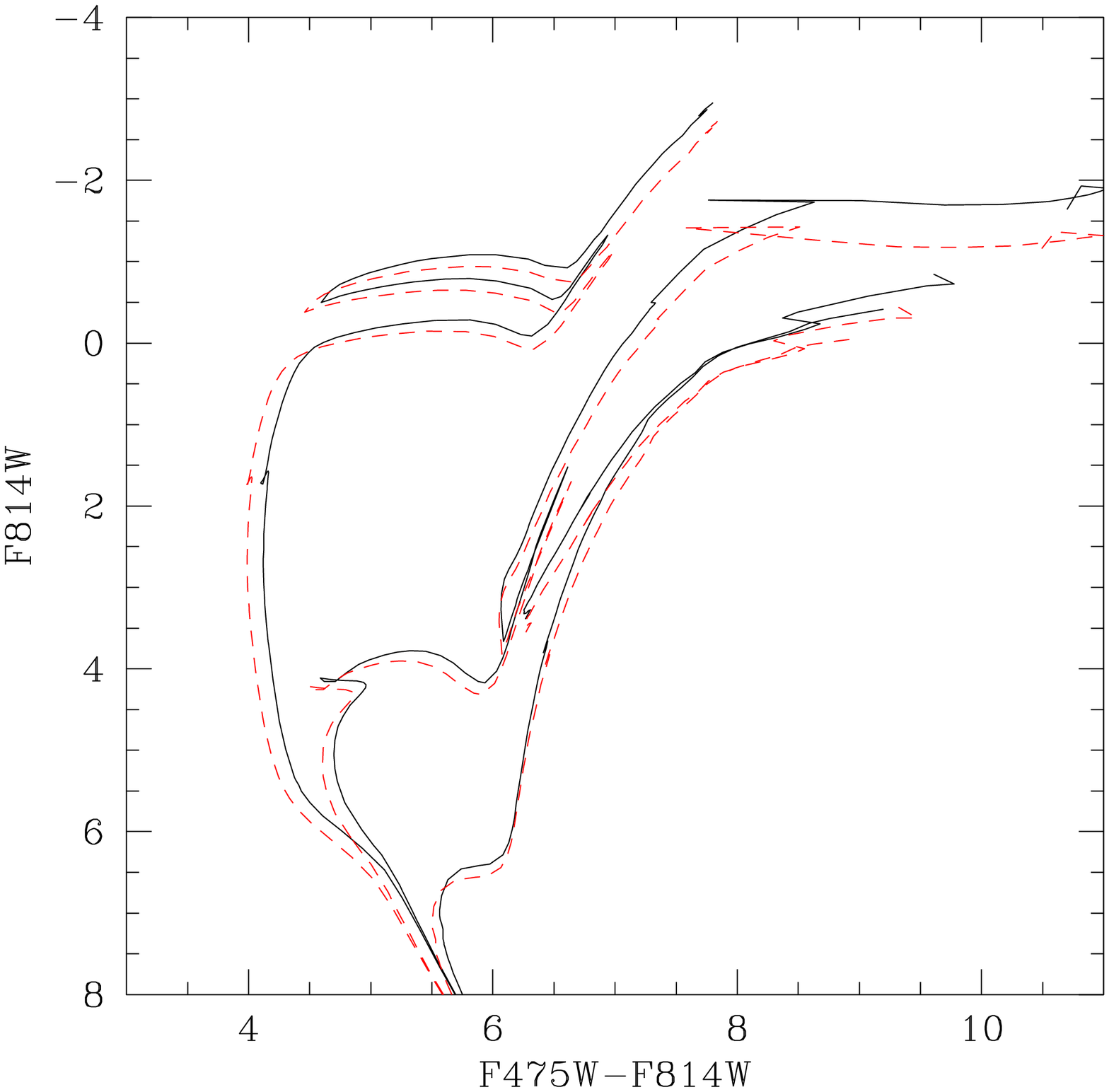}
\epsscale{1.0}
\caption{A set of $Z=0.008$ isochrones from Marigo et al. (2008) in 
the ACS/WFC system, extincted by $A_V=6$~mag according to two
different prescriptions: adopting the star-to-star interstellar
extinction coefficients as described in this paper (continuous lines)
and adopting the coefficients derived from a yellow dwarf at the
regime of low interstellar extinction (dashed lines). The isochrone
ages are, from top-left to bottom-right, 0.1, 1, and 10 Gyr.}
\label{fig_isocav}
\end{figure}

\placefigure{fig_isocav}

Figure \ref{fig_isocav} presents a practical application of the
star-to-star interstellar extinction coefficients derived in this
work, as compared to the usual approximation of applying a single set
of coefficients for all stars and independently of the actual amount
of total extinction. In the figure, one can notice that a set of
isochrones extincted by $A_V=6$~mag in the correct (star-to-star) way,
is narrower in colour (by about 0.3~mag in F475W--F84W) than an
isochrone extincted by the same $A_V$ but using a single set of
coefficients derived from a low-extinction yellow dwarf. Moreover,
using the right interstellar extinction coefficients produces brighter
red giants than in the case of constant coefficients. This latter
effect is striking in the comparison between the AGB sequences at the
top-right corner of Fig.~\ref{fig_isocav}.

Although the $A_V=6$~mag case is rather extreme, it well illustrates
the effect of star-to-star interstellar extinction. Whether this
effect is important depends on the photometric accuracy of the data
and models available, and on the passbands being considered. A very
approximate rule is that, if one aims at reproducing data with an
accuracy better than 0.05~mag in optical bands, star-to-star
interstellar extinction should start being considered already at
$A_V>2$~mag. A more consistent and safer rule, however, is to {\em
always} take star-to-star extinction into consideration.

\section{Application to isochrones and data release}
\label{sec_isoc}

Extensive tables with BC and interstellar extinction coefficients have
been computed and are provided in the static web repository
\url{http://stev.oapd.inaf.it/dustyAGB07}. The BC tables contain 
entries for all stars in our spectral library. The extinction
coefficients instead have been computed for solar metallicity only. In
fact, the effective temperature, and to a lower extent $\log g$, is
the main parameter driving the changes in $A_{S_\lambda}/A_V$. We have
verified that $A_{S_\lambda}/A_V$ values change very little as a
function of [M/H]: as an example, for a dwarf star of $\Teff=4000$~K
extincted by $A_V=1$, $A_{\rm F475W}/A_V$ values differ by just
0.17~\% between the $\mh=0$ and $\mh=-2.5$ cases.

Moreover, the interactive web interface
\url{http://stev.oapd.inaf.it/cmd} allows the present tables to be
applied to the theoretical isochrones from Padova that are based on
Girardi et al. (2000) tracks. Taking into account the Bertelli et
al. (1994) isochrone models for initial masses higher than 7~\Msun,
interpolated isochrones can be constructed for any age between 0 and
17 Gyr, and for any metallicity value between $Z=0.0001$ and $Z=0.03$
(see Girardi et al. 2002). A recent paper by Marigo et al. (2008)
presents a substantial revision of these isochrones, with the
replacement of the old TP-AGB tracks by much more detailed ones
(cf. Marigo \& Girardi 2007). The same isochrones will soon be
extended to the planetary nebulae and white dwarfs stages.

A novelty in this latter web interface is the possibility of producing
isochrones already incorporating the effect of interstellar extinction
and reddening. This is done via interpolations in the tables of
$BC_{S_\lambda}(A_V)$, using $\log \Teff$, $\log g$, and $A_V$ as the
independent parameters.  The dependence in metallicity is ignored
since it is really minimal (with $\la0.2$~\% differences in
$A_{S_\lambda}/A_V$, going from the solar-metallicity to the
very-metal poor cases). Therefore, the $BC_{S_\lambda}(A_V)$ tables
are prepared for $\mh=0$ only, and the
$BC_{S_\lambda}(A_V)-BC_{S_\lambda}(0)$ corrections derived from these
tables are applied at all metallicities.  This conceptually simple
procedure is completely equivalent to alternative ones based on the
interpolation of $A_{S_\lambda}/A_V$ tables.

In this way, users of the web interface just need to specify the total
interstellar extinction $A_V$, to apply the proper $A_{S_\lambda}$
value at every point in the isochrone in each filter passband. We
remind that the traditional approach is to apply a single value of
extinction $A_{S_{\lambda_1}}$ and reddening
$A_{S_{\lambda_1}}-A_{S_{\lambda_2}}$ to all points along an isochrone
-- $A_{S_{\lambda_1}}$ and $A_{S_{\lambda_2}}$ being related to the
total $A_V$ via constant multiplicative factors -- then occasionally
seeking for the $A_V$ value that best fits a certain kind of
observations. Instead, by applying the proper $A_{S_\lambda}$ value to
each isochrone point, we take into account its dependence on
\Teff, on $\log g$, and on the total $A_V$ itself. Our procedure will
lead to both (1) the traditional overall translation of isochrones in
the CMD, and (2) moderate changes in the isochrone shapes as
interstellar extinction increases, especially for colours involving
blue passbands and in the regime of high extinction. The changes in
the isochrone shapes will not be noticeable for low extinctions (for
$A_V\la2$~mag, in optical passbands) and/or for near-infrared
passbands.

Moreover, the present tables of bolometric corrections, for the
no-extinction case only, are being applied to Bertelli et al. (2008)
isochrones, which cover a wide region of the helium--metal content
$(Y,Z)$ plane.

\section{A few comparisons with data}
\label{sec_data}

The present bolometric corrections and interstellar extinction
coefficients for ACS and WFPC2 are part of a large database of
transformations to many photometric systems (see Marigo et al. 2008
for the most updated version). Regarding the BC and colour-\Teff\
relations for non-C stars, they have just slightly changed -- a few
hundredths of magnitude at most -- with respect to those presented by
Girardi et al. (2002). Therefore the transformations presented in this
paper are in fact already largely tested in the literature, in many
papers that compared data and models seeking accuracies to within
$\sim0.05$~mag in colours. Most of these previous comparisons were
however done in Johnson-Cousins, 2MASS and SDSS systems.

In the present paper we limit ourselves to a couple of simple
comparisons between present models and observations in the ACS/WFC
F475W, F606W and F814W pass-bands. These filters, together with F555W,
are those in which most observations of resolved galaxies are
available. These are also the filters used by the ANGST project.

\begin{figure}[h]
\plottwo{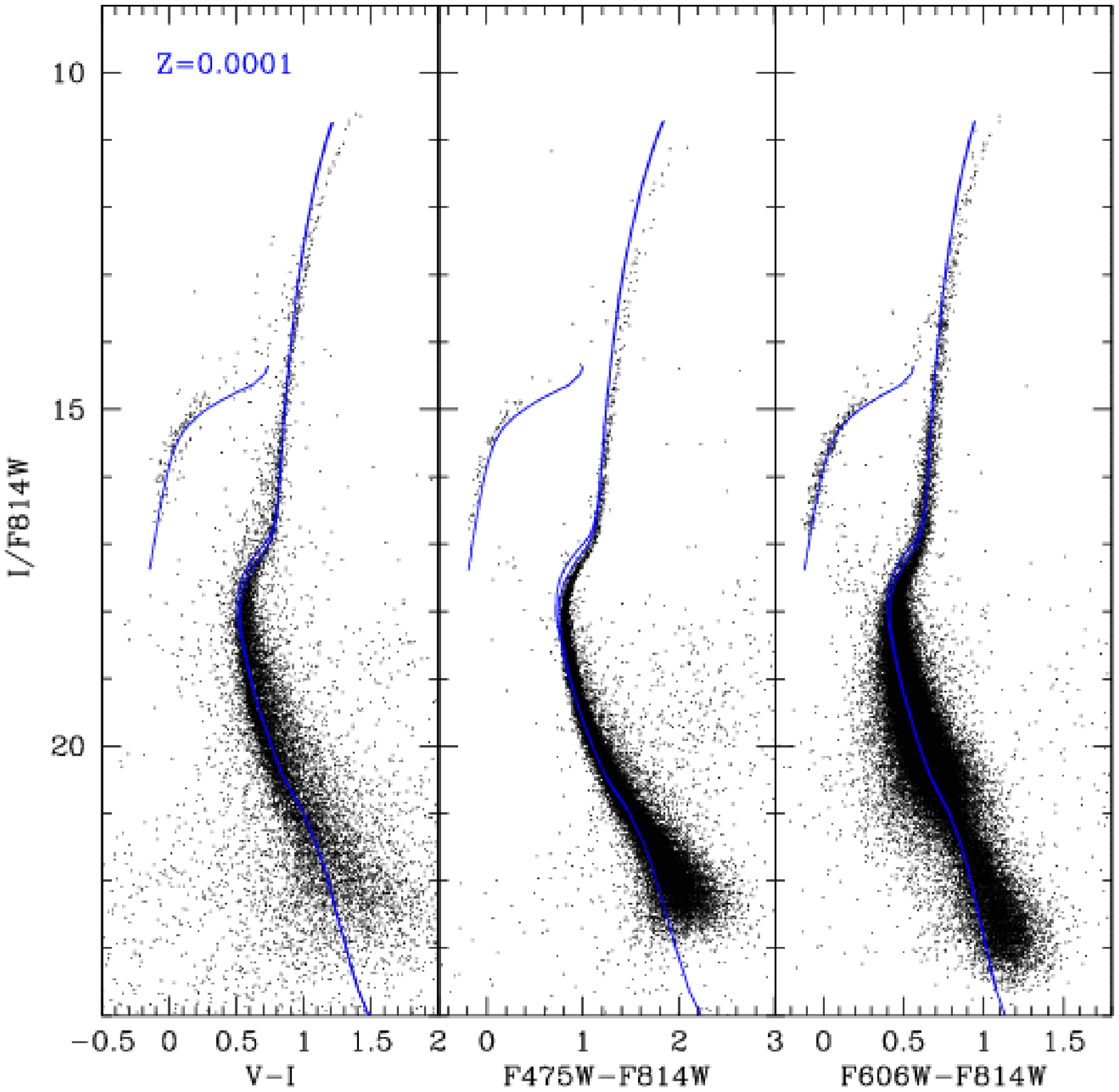}{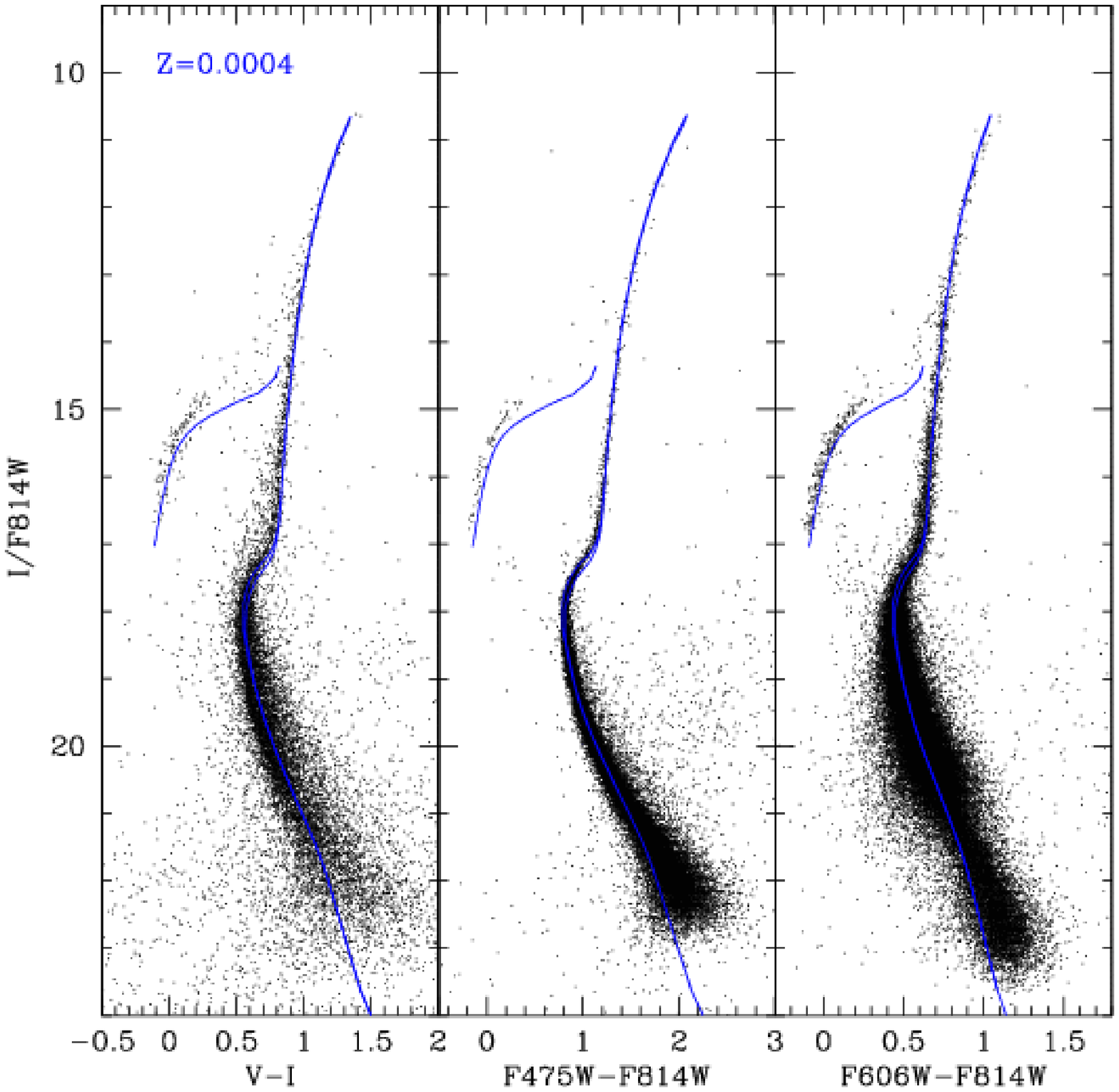}
\caption{CMD of M92 in three different band combinations, compared with 
isochrones of metallicity $Z=0.0001$ and $0.0004$ as indicated in the
labels. The isochrone ages are log(age/yr)=10.10 and 10.15. See the
text for details.}
\label{fig_data1}
\end{figure}

The two panels of Fig.~\ref{fig_data1} compare $Z=0.0001$ and
$Z=0.0004$ isochrones superimposed to the CMD of the globular cluster
M92 in three band combinations: $(V-I)$, (F475W--F814W) and
(F606W--F814W). The $VI$ data are from the Rosenberg et al. (2000)
database, and has kindly been provided by A. Rosenberg. The F475W,
F814W CMD has been obtained using archival data from the ACS LCID
project\footnote{The LCID (Local Cosmology from Isolated Dwarfs
project, HST P.ID. 10505 and 10590) has obtained deep CMDs reaching
the oldest main sequence turnoffs in five isolated Local Group
galaxies and two clusters, M92 and NGC 1851.} and the F606W, F814W CMD
using ACS archival data of P.ID. 9453 (see Brown et al. 2005). The
isochrones have log(age/yr)=10.10 and 10.15, and are plotted from the
lower main sequence up to the TRGB; additionally, we plot the zero-age
horizontal branch from Girardi et al. (2000) tracks as derived by
Palmieri et al. (2002).  In all cases we have assumed a distance
modulus $(m-M)_0=14.62$ (del Principe et al. 2005), and interstellar
extinction $A_V=0.068$ following the Schlegel et al. (1998) maps. The
cluster measured metallicity is $\feh=-2.24$ (Zinn \& West 1984); or
$-2.16$ (Carretta \& Gratton 1997), which would correspond to
$Z=0.0001$ of Girardi et al. (2000) scaled-solar isochrones. This case
is plotted in the upper panel. Note however that the $Z=0.0004$
isochrones (lower panel) seem to fit the cluster sequences best,
especially the RGB slope and the position of the main sequence
turn-off. Note that the different mismatches that can be appreciated
go in the same direction in the ground-based and the ACS fit. These
mismatches are likely to be associated to the theoretical isochrones
themselves, rather than to the color transformations.

\placefigure{fig_data1}

\begin{figure}[h]
\epsscale{0.7}
\plotone{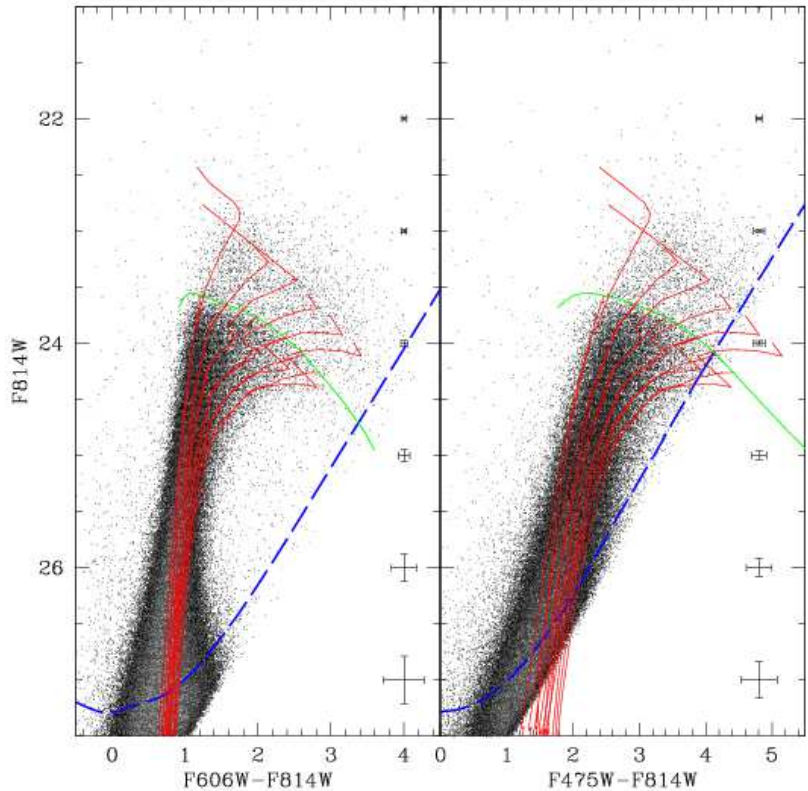}
\epsscale{1.0}
\caption{CMDs for the ACS Wide~1 pointing over NGC~253 (black and gray 
dots). The dashed line signals the 50~\% completness limit of the ACS
observations. The red lines are Marigo et al. (2008) 10-Gyr isochrones
of metallicities $\mh=(-2.0, -1.5, -1.0, -0.5, 0.0)$, shifted by a
distance modulus of 27.6. The green line marks the predicted locus of
the TRGB for a more complete sequence of 10-Gyr isochrones with \mh\
increasing from $-2.3$ to $+0.1$.}
\label{fig_data2}
\end{figure}

This comparison with a nearby globular cluster illustrates the
usefulness of the isochrones in the interpretation of star clusters in
general, in which the photometry goes from the red giants down to the
unevolved section of the main sequence. Most of the HST archival data
refers instead to the composite stellar populations of nearby
galaxies. Fig.~\ref{fig_data2} shows the case of field WIDE1 located
about 13~arcmin along the major axis of the spiral galaxy NGC~253,
observed with ACS/WFC for the ANGST project. Surprisingly, this region
shows no sign of young stellar populations. A set of 10-Gyr old
isochrones overimposed on the data CMDs would indicate a range of \mh\
going from about $-2$ to $0$. The observed CMD presents a striking
decrease of stellar density above a certain level of the red giants,
which obviously corresponds to the TRGB position. In
Fig.~\ref{fig_data2}, the green line marks the TRGB for a series of
10-Gyr-old isochrones of varying \feh, displaced by a distance modulus
of $27.6$~mag. It can be noticed that this line describes remarkably
well the shape of observed boundary between the stellar-rich RGB and
the less populated AGB. Although the precise position of this TRGB
line is determined also by the quality of the underlying stellar
models, its reproduction over a wide range in colour (covering more
than $2$~mag in F475W--F814W, and more than $1$~mag in F606W--F814W)
is clearly suggesting a correct behaviour of the set of BCs that has
been applied.

\placefigure{fig_data2}

\section{Concluding remarks}
\label{sec_future}

In this paper we present tables of bolometric corrections and
interstellar extinction coefficients, as a function of \Teff, $\log
g$, and [M/H], for the WFPC2 and ACS systems. We keep the original
definition of the Vega magnitudes by Holtzman et al. (1995) and
Sirianni et al. (2005), while adopting the latest accurate
measurements of the Vega fluxes by Bohlin (2007). The tables are
applied to theoretical isochrones in
\url{http://stev.oapd.inaf.it/cmd} and
\url{http://stev.oapd.inaf.it/dustyAGB}, where similar data for many 
other photometric systems can be found. A major novelty of this work
is that interstellar extinction coefficients are computed and applyed
star-by-star, so that isochrones present not only the overall shift in
the CMDs as extinction increases, but also the subtle changes in their
shape. We have made simple comparisons of these isochrones with real
data, but a more careful testing of the transformations is left to
future papers, and to the interested readers.

To the users of these HST isochrones, suffice it to recall a few
general caveats, which apply to all photometric systems: First, our
results are based on theoretical model atmospheres and spectral energy
distributions, which are known not to be accurate at the
hundredths-of-magnitude level, especially for the coolest stars. The
same applies to Bohlin's (2007) Vega spectrum, which has uncertainties
of 1--2~\% in the optical. Moreover, the results for narrow-band
filters are particularly sensitive to errors in line opacity lists,
and hence may be more uncertain. Overall, we expect the BCs for
broad-band filters to be accurate to within $\sim0.05$, with errors in
colours being somewhat smaller. If we consider all the other
uncertainties involved in the analysis of HST photometric data --
including for instance the substantial uncertainties in the
metallicity distribution and interstellar extinction curves of nearby
galaxies -- errors of this magnitude may be considered as acceptable.

The data provided in this paper are not conceived as a static set of
tables, but rather as a database that will be revised and/or extended
as soon as there is a significant improvement on the input ingredients
-- comprising the spectral flux libraries, the reference Vega
spectrum, filter transmission curves, the interstellar extinction
curve, and the isochrone sets to which the transformations are
applied. An important extension is the inclusion of tables for the
WFC3 camera (Bond et al. 2006), which will soon replace WFPC2 after
HST Servicing Mission 4. Preliminary tables for WFC3 are already
provided in the present database since they may be useful in the
preparation for HST Cycle 17. They were derived in the same way as
WFPC2 and ACS tables, using filter throughputs from SYNPHOT/STSDAS
v3.7 files available at January 15, 2008. The WFC3 data will be
replaced as soon as revised throughputs are released by the WFC3 team.

\acknowledgments

We thank M. Sirianni, B. Salasnich and B. Balick for their
early interest in these isochrones and for their help with the filter
transmission curves.  L.G. acknowledges A. Bressan and P. Marigo for
their encouragement, and the many people who have pointed out problems
in previous versions of the ACS and WFPC2 isochrones.



{\it Facilities:}  \facility{HST (ACS)}, \facility{HST (WFPC2)}.





\end{document}